\documentclass[preprint,showpacs,preprintnumbers,amsmath,amssymb,superscriptaddress]{revtex4}
\usepackage{graphicx}
\usepackage{dcolumn}
\usepackage{bm}

\begin{document}

\title{Recombination processes in CuInS$_{2}$/ZnS Nanocrystals during steady-state photoluminescence}


\author{Yue Sun}
\author{Chenjiang Qian}
\author{Kai Peng}
\affiliation{Beijing National Laboratory for Condensed Matter
Physics, Institute of Physics, Chinese Academy of Sciences, Beijing,
100190, China}

\author{Zelong Bai}
\affiliation{Beijing Key Laboratory of Nanophotonics and Ultrafine Optoelectronic Systems, School of Materials Science $\&$
Engineering, Beijing Institute of Technology, Beijing, 100081, China}
\author{Jing Tang}
\author{Yanhui Zhao}
\author{Shiyao Wu}
\author{Hassan Ali}
\author{Feilong Song}
\affiliation{Beijing National Laboratory for Condensed Matter
Physics, Institute of Physics, Chinese Academy of Sciences, Beijing,
100190, China}

\author{Haizheng Zhong}
\affiliation{Beijing Key Laboratory of Nanophotonics and Ultrafine Optoelectronic Systems, School of Materials Science $\&$
Engineering, Beijing Institute of Technology, Beijing, 100081, China}
\author{Xiulai Xu}
\email{xlxu@iphy.ac.cn}
\affiliation{Beijing National Laboratory for
Condensed Matter Physics, Institute of Physics, Chinese Academy of
Sciences, Beijing, 100190, China}

\begin{abstract}

We report on temperature- and excitation-power-dependent photoluminescence (PL) study of CuInS$_{2}$/ZnS nanocrystals dispersed on a SiO$_{2}$/Si substrate with a confocal micro-PL system. With increasing the excitation power at 22 K and room temperature, the PL spectra are blue-shifted because of the state filling. At low temperature, a small peak is observed at the low energy side of the spectrum, which could be due to the F$\ddot{o}$rster resonance energy transfer between different nanocrystals. The integrated PL intensity increases sublinearly as a function of excitation power with a power factor of around 2/3, which demonstrates the Auger recombination dominated process in the nanocrystals, especially under the high excitation power.

\end{abstract}
\maketitle

Colloidal semiconductor nanocrystals (NCs) also called quantum dots (QDs) have attracted great attentions because
of their potential applications on light emitting diodes (LEDs), \cite{Shirasaki2013} solar cells,\cite{Tang2011,Chuang2014} lasers,\cite{Wang2014,Park2015,Yue2015,Jun2015} bio-imaging\cite{Kershaw2013} and optical quantum information processing.\cite{Moro2014,Xu07,Xu08,Mar2014} Up to now, controlling over the composition, size, shape, crystal structure, and surface properties of colloidal nanocrystals is well explored to enhance the quantum yield (QY) for these applications.\cite{Kovalenko2015} For example, near unity QY has been demonstrated with colloidal nanocrystals lately.\cite{Nasilowski2015,Xingliang2014} Due to the high yield, single CdSe colloidal nanocrystals have been investigated to achieve single photon emission for quantum information processing.\cite{Michler2000,Brokmann2004,Wissert2011} Additionally, the harvesting solar application of colloidal NCs has been intensively investigated and the power conversion efficiency approaches 8$\sim9$\%.\cite{Chuang2014,Zhao2015} LEDs based on the colloidal NCs have been demonstrated with multi-color emission by controlling the sizes and the components.\cite{Shirasaki2013,Yang2015,Caruge2008} However, the above progresses were focused on Lead or Cadmium based compounds, it is much aspired to explore less toxic compounds for future commercialization.

A new type of colloidal NCs belonging to \uppercase\expandafter{\romannumeral1}-\uppercase\expandafter{\romannumeral3}-\uppercase\expandafter{\romannumeral6} group have been synthesized lately, such as CuInS$_2$ (CIS) NCs and CuInSe$_2$ NCs, which have low toxicity and tunable emission wavelength from visible to near infrared region while \uppercase\expandafter{\romannumeral2}-\uppercase\expandafter{\romannumeral6} group NCs contain toxic elements. Chen \emph{et al}. \cite{Chen2012,Chen2013} have demonstrated white light emitting diodes with CIS nanocrystals as phosphors for red emission. Up to now, the radiative decay of CIS NCs has generally been attributed to defect states.\cite{Zhong2012,Shi2012} However, the exact recombination mechanism is still under debate with defect states, such as recombination from conduction band to localized intra-gap state,\cite{Li2011} donor-acceptor pair (DAP),\cite{Zhong2012,Shi2012} from a localized state to valence band.\cite{Omata2014} Therefore, it is worthwhile to understand the luminescence mechanism of CIS NCs as potential materials for LEDs and solar cells. Comparing with II-VI materials, CIS NCs have a lower QY which has been attributed to the defect induced non-radiative recombination.\cite{Zhong2012} Additionally during the recombination, the non-radiative processes such as the trap-mediated Shockley-Read-Hall (SRH) recombination and Auger recombination in colloidal quantum dots could also play an important role.\cite{Seetoh2013,Jang2008,Zabel2015}  Auger process is known that the energy of the electron and hole recombination is transferred to a third carrier, and the carrier either an electron or a hole is reexcitated to higher energy state, resulting in no light emission.\cite{Htoon2003}

In this Letter, we study the steady-state PL of the Cu-deficient CIS/ZnS core/shell NCs with the [Cu]/[In] ratio of 0.25 with changing the excitation power and the sample temperature. Due to the size distribution of NCs, the PL spectra can be well fitted with Gaussian shape. A small peak can be observed at the low energy side of the spectrum at low temperature, which could be due to the F$\ddot{o}$rster resonant energy transfer (FRET). At room temperature, the peak energy of NCs shows a  blueshift of about 28 meV with increasing excitation power because of the state filling.\cite{Pan2015} An Auger recombination process is demonstrated in the CIS/ZnS core/shell NCs both at 300 K and 22 K.\cite{Seetoh2013}

CIS nanocrystals were synthesized by thermolysis of a nontoxic mixture solution of Copper (I) acetate, Indium acetate and alkylthiol in a high boiling point solvent at 240 $^{\circ}$C. The details of the synthesis can be found in Ref.\cite{Zhong2008,Chen2012}. After the synthesis, the CIS nanocrystal powder was dissolved in alcohol with different concentrations. The dispersed solutions were dropped on a SiO$_{2}$/Si substrate and the close-packed CIS NC solids were formed after the alcohol evaporated.\cite{Pan2015} With this method, we prepared three densities of close-packed NC solids labeled as NC1, NC2 and NC3 (NC1:NC2:NC3=5:3:2) with relative density from high to low in the solutions. A typical scanning electron microscope (SEM) image of the nanocrystals on the substrate of NC1 is shown in the inset of Figure 1 (a), which shows a good uniformity. The dots in the SEM image are CIS/ZnS NCs, and the size of NCs obtained from high-resolution transmission emission microscopy (HRTEM) in Ref.\cite{Chen2012} is around 5 nm. The typical sizes of the nanocrystals are around 2-5 nm with emission wavelengths varying from 525 nm (green) to 750 nm (near infrared).\cite{Chen2012,Chen2013} The samples were mounted on a cold-finger cryostat in vacuum, with which the sample temperature can be tuned from 22 K to room temperature. The PL measurement was performed by a conventional confocal micro-PL system. A HeCd laser with a wavelength of 325 nm was used as an excitation source and was focused on the sample with a spot size of 1-2 $\mu$m in diameter by a large numerical aperture objective. The emitted light was collected with the same objective and then dispersed through a 0.55 m spectrometer. The spectrum was detected with a charge coupled device camera cooled with liquid nitrogen.

Figure 1 (a) shows a typical PL spectrum of NC1 with an excitation power about 5.32 $\mu$W, which comes from the emission of the CIS/ZnS core/shell dots. It can be well fitted with a Gaussian shape with a peak energy at 2.048 eV as shown by red-dashed line, indicating that the spectrum broadening is due to the inhomogeneous size distribution of NCs. When the temperature goes down to 22 K, the spectrum can not be well fitted by a single Gaussian peak as shown by the top traces in Figure 1 (b), especially at the low energy tail of the PL spectrum. So the PL spctrum at low temperature was fitted by two Gaussian peaks, which are represented by the bottom traces in Figure 1 (b). A small peak at 1.68 eV at the low energy side (red-dashed line) can be resolved in addition to the main PL peak (green-dashed line) at 2.015 eV. Considering the distance less than 10 nm between different nanocrystals (as shown in the inset of Figure 1(a)), the small peak could be attributed to the FRET from small quantum dots to large dots because of the spectrum overlapping between the two peaks as shown in the Figure.\cite{Valerini2005,Li2012}

The integrated PL intensities of the small peak at 1.68 eV as a function of temperature are shown in Figure 1 (c) with excitation power at 0.07, 0.14, 1.4 and 7.0 $\mu$W. The PL intensity increases with the increasing pumping power, which means that the FRET process is enhanced with increasing photogenerated carrier density. Kagan \emph{et al}.\cite{Kagan1996} reported that the electronic energy transfer from small NCs to large NCs can be enhanced by thermal energy. However in Figure 1 (c), the small peak intensity decreases generally as a function of temperature. With increasing the sample temperature, the linewidth of main peak increases because of the phonon-assisted broadening (as shown in Figure 2 (c)). The higher temperature the broader linewidth of the main peak, which induces that the whole spectrum can be well fitted with a single Gaussian peak in the end. As a result, the small peak intensity decreases with increasing temperature and finally cannot be resolved.

To further understand the recombination mechanism of the NCs, we only consider the main peak in the PL spectra in the rest of this work by single-peak fitting at 300 K and two-peak fitting at 22 K respectively. Through altering the excitation power, we investigated the evolution of PL spectra of the close-packed CIS/ZnS NC solids at room temperature as shown in Figure 2 (a). The PL intensity increases quickly with increasing excitation power and saturates. Blueshift is observed as well. In order to display the blueshift clearly, the PL spectra pumped by different excitation power are normalized as shown in Figure 2 (b). An energy increase of about 20 meV (from 2.03 eV to 2.05 eV) can be resolved with increasing the excitation power from 0.885 $\mu$W to 14.63 $\mu$W. The power-dependent peak position and linewidth are summarized in Figure 2 (c) at 300 K and 22 K. With the increasing excitation power, both the peak energies exhibit large blueshifts of about 28 meV (300 K) and 8.5 meV (22 K) and have a tendency to saturate, which can be ascribed to the state filling. At 300 K, a strong linewidth broadening can be observed, with a full width at half maximum (FWHM) changing from 0.37 eV to 0.40 eV within the pumping power range. However at 22 K, the change in FWHM is not obvious as a function of the excitation power. We attribute the linewidth broadening at room temperature to the phonon-assisted processes. It should be noted, the peak energy at room temperature is higher than that at 22 K. The optically generated carriers at low energy levels could be thermally activated to high levels with the temperature increasing, which can induce the blueshift, and after getting rid of the thermal activation, the carriers would hop from high energy levels to low levels.\cite{Eliseev1997}

In general, the mechanism of photoluminescence consists of radiative and non-radiative recombinations. The radiative recombinations of nanocrystals are mainly due to the exciton recombination in CdSe and SnO$_{2}$ quantum dots,\cite{Nasilowski2015,Valerini2005,Pan2015} or the defect-related recombination in CIS NCs.\cite{Zhong2012,Shi2012,Li2011,Omata2014} In contrast, the non-radiative processes including the trap-mediated SRH recombination and Auger recombination could also dominate the relaxation process in colloidal quantum dots, especially in CIS NCs. Here, we consider that the integrated PL intensity \emph{I} depending on the excitation power \emph{P} can be expressed as power law $I\varpropto P^{k}$, where \emph{k} is a constant, reflecting various recombination processes.\cite{Seetoh2013}  If \emph{k}$>$1, the SRH recombination also called trap-assisted recombination will occur. If \emph{k}=1, it means that radiative recombination is the main process. If \emph{k}$=$2/3, the Auger recombination is dominant, especially in high carrier density. The pumping power \emph{P} can be expressed with these recombinations as \cite{Seetoh2013}
\begin{eqnarray}
\alpha P=R_{L}+R_{Aug}+R_{SRH},
\end{eqnarray}
where $\alpha$ is a constant, and $R_{L}$, $R_{Aug}$ and $R_{SRH}$ represent the rates of radiative recombination, Auger recombination and trap-mediated SRH recombination, respectively.
With
\begin{eqnarray}
 R_{L}=\beta I\propto P^{k},
 \end{eqnarray}
where $\beta$ is a constant. Then, the $R_{L}$, $R_{Aug}$ and $R_{SRH}$  can be written as
\begin{eqnarray}
 R_{L}=B_{L}(n_{0}+\Delta n)\Delta n,
\end{eqnarray}
\begin{eqnarray}
R_{Aug}=B_{Aug}(n_{0}+\Delta n)^{2}\Delta n,
\end{eqnarray}
\begin{eqnarray}
R_{SRH}=B_{SRH}\Delta n,
\end{eqnarray}
where $ B_{L}$, $B_{Aug}$, $B_{SRH}$ represent radiative recombination, Auger recombination and SRH recombination coefficient, respectively. $\Delta n$ is the photogenerated electron density at equilibrium and $ n_{0}$ is the intrinsic electron density. In PL of CIS NCs, the peak energy with increasing excitation power displays a blue-shift by state filling (as shown in Figure 2 (c)), which is induced by the photogenerated carriers. This result can fulfill the condition that $\Delta n \gg n_{0}$. When $\Delta n\gg n_{0}$, we can calculate that $R_{L}\propto \Delta n^{2}$ and $R_{Aug}\propto \Delta n^{3}$, where $I\propto R_{L}\propto \Delta n^{2} $. Then assuming that the dominant luminescence mechanism is Auger recombination, the excitation power $P\propto R_{Aug}\propto \Delta n^{3} $, eventually obtaining $I\propto P^{2/3}$. When the SRH recombination is taken over, the result could be $I\propto P^{2}$. Similarly, when the radiative recombination is the main transition process, the value of \emph{k} is equal to 1.

The PL spectra were collected with different excitation power for three samples (NC1, NC2 and NC3). The integrated intensities of the PL spectra for each sample as a function of the excitation power are illustrated in Figure 3 (a) and (b) for sample temperature at 300 K and 22 K respectively. It can be seen that the PL intensity increases linearly with increasing excitation power in logarithmic coordinate especially in the high power regime. We use the power law to fit the data from Figure 3 and obtain the value of \emph{k} of each sample as shown in the inset. The value of \emph{k} close to 2/3 demonstrates that the Auger effect plays a crucial role in the luminescence mechanism, even at room temperature. At the low excitation power regime, the integrated PL intensities deviated from the fitted lines at both temperatures with a larger \emph{k}. This reveals that relaxation mechanism is mainly radiative recombination at low excitation power, while the Auger effect is much weaker. Normally the radiative recombination could originate from the exciton recombination.\cite{Schmidt1992} However in the CIS quantum dots, a large Stokes shift has been observed, which implies that the emission comes from the defect states instead of exciton transition.\cite{Bose2015} Furthermore, the emission peak have a blueshift with the increasing excitation power in Figure 2 (c), but for free excitons, the position of the energy peak should be independent of the excitation power without considering the heating effect.\cite{Castro2004} Therefore, it is widely considered that the radiative recombination comes from the defect states.\cite{Zhong2012,Shi2012,Li2011} The fitted values of \emph{k} with low excitation power at 22 K (as shown in the inset of Figure 3 (b)) are around 1, which also further confirms that recombination is defected related at low excitation power regime.\cite{Schmidt1992}

In summary, we have shown the PL spectra evolution of the close-packed CIS/ZnS core/shell nanocrystals on SiO$_{2}$/Si substrates with changing the excitation power and temperature. A small peak was observed at 1.68 eV at 22 K, which is due to the F$\ddot{o}$rster resonant energy transfer from the small dots to large dots. The peak energy of nanocrystals has a pronounced blueshift both at room temperature and low temperature because of the state filling. With increasing excitation power, a linewidth broadening of about 30 meV is observed at room temperature but only slight broadening happens at 22 K, indicating that the linewidth broadening is assisted by the phonons. By fitting the integrated PL intensity as a function of excitation power, the power factors of three samples around 2/3 reveal that the Auger recombination dominates at room temperature and 22 K, especially at high excitation power. The evidence of the Auger recombination process in CIS NCs is useful to improve the quantum yield for future applications in colloidal quantum dot based optoelectronic devices.

\begin{acknowledgments}
This work was supported by the National Basic
Research Program of China under Grant No.
2013CB328706 and 2014CB921003; the National
Natural Science Foundation of China under Grant No. 91436101, 11174356 and 61275060; the
Strategic Priority Research Program of the
Chinese Academy of Sciences under Grant No.
XDB07030200; and the Hundred Talents Program
of the Chinese Academy of Sciences.
\end{acknowledgments}

\newpage
\begin{figure}[htp]
\centering
\includegraphics[scale=0.7]{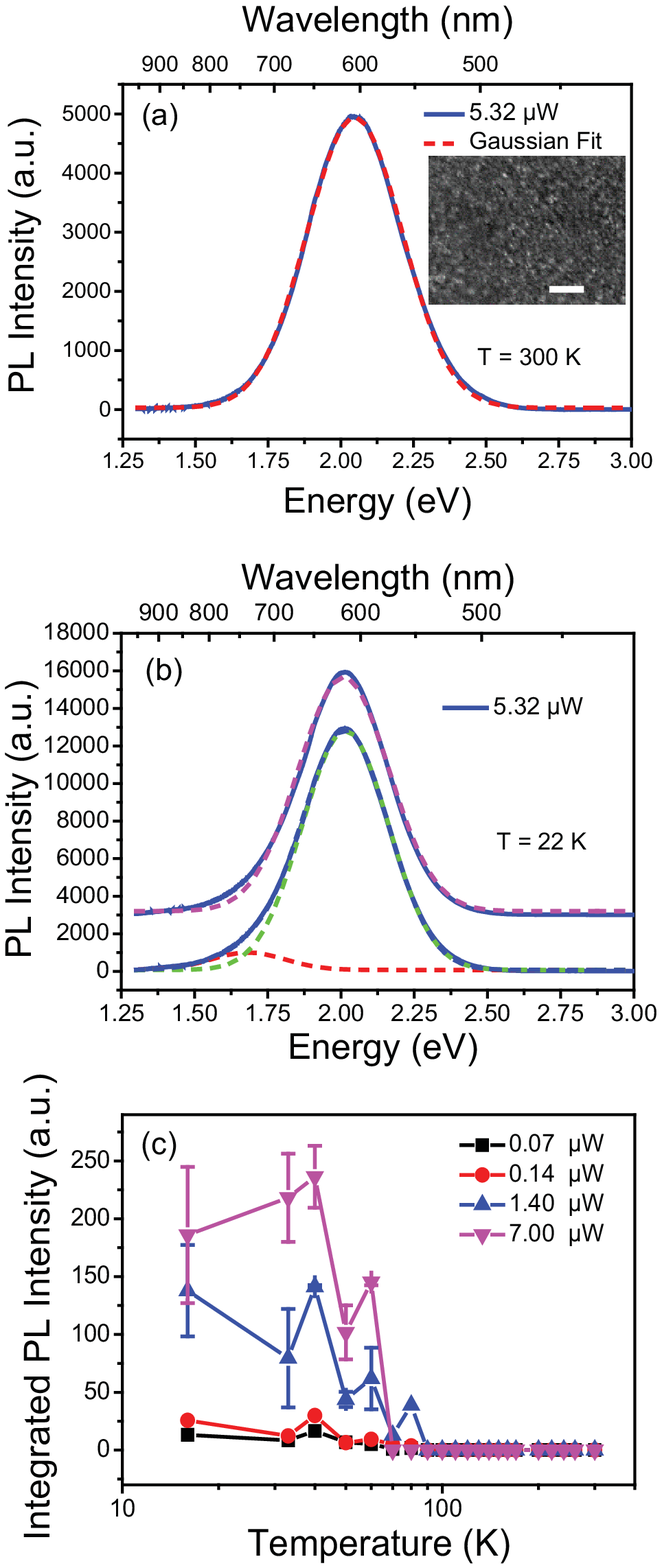}
\caption{(a) A typical PL spectrum of close-packed CIS/ZnS NC solids obtained at room temperature with an excitation power of 5.23 $\mu$W, and the red-dashed line shows the fitted result by a Gaussian shape with a peak energy at 2.048 eV. The inset shows a SEM image of sample NC1 and the white bar is 100 nm. (b) The blue lines are the PL spectra at 22 K, the dashed lines are fitted results. The two set of spectra are shifted for clarity. (c) The integrated PL intensity of the low energy peak as a function of temperature with different excitation power. The large error bars are due to the peak position change of the small peak during the fitting.}
\label{Confocal}
\end{figure}

\newpage
\begin{figure}[htp]
\centering
\includegraphics[scale=0.7]{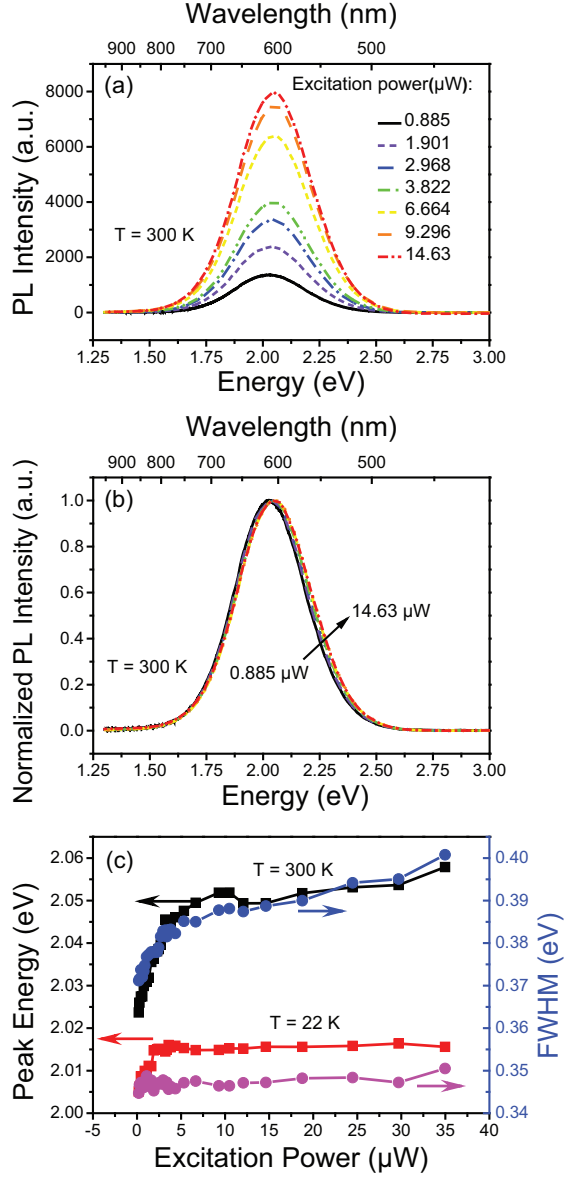}
\caption{(a) The evolution of PL spectra of CIS/ZnS NCs at different excitation power under 300 K. (b) Normalized PL spectra of NCs showing a blueshift from 0.885 $\mu$W to 14.63 $\mu$W. (c) Peak energy and FWHM as a function of excitation power under 300 K and 22 K. The black and red squares represent the peak energy, which show the blueshift because of the state filling. FWHMs are shown by blue and pink circles.}
\label{Confocal}
\end{figure}

\newpage
\begin{figure}[htp]
\centering
\includegraphics[scale=0.7]{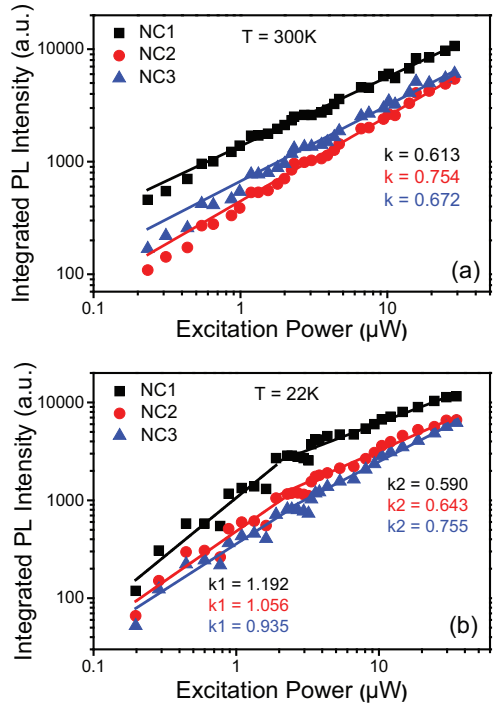}
\caption{The integrated PL intensity of three samples with different concentration as a function of excitation power at 300 K (a) and 22 K (b). The inset shows the fitted values of \emph{k} of three samples at both temperatures. }
\label{Confocal}
\end{figure}

%

\end{document}